# The Impacts of Using Business Information Systems on Operational Effectiveness in Hungary

*Péter Sasvári*
*Institute of Business Sciences,*
*University of Miskolc, Hungary*

*Abstract— Business expectations regarding the introduction of business information systems were investigated according to company size categories. The results clearly showed that according to the majority of the respondents the information supply for decision-makers improved. In contrast, business information systems as a means of improving competitiveness were only regarded by corporations, this aspect was only around the average in the other company size categories. The respondents evaluated to what extent the usage of business information system provided assistance for their economic analyses. The obtained results show that business information systems can be utilized well in controlling and reporting. There are differences in their judgement by size categories. Particularly corporations can take advantage of the support of business information systems mainly in the field of planning, plan-actual analysis and the exploration of cost reducing possibilities.*

## I. INTRODUCTION

The role of information has become more and more substantial in the economy recently, and information is regarded as an important resource since it is more difficult for companies to improve their market positions in the long term without having the appropriate amount of available information [2][3]. Globalization in the business world has brought about the possibility of getting a greater amount of information in much less time which means that companies are forced to spend more time and energy on handling the increased information load [5] [6] [12]. Business information systems are designed to provide effective help in this process as they are becoming increasingly popular among companies due to the robust technological development. This paper deals with the usage of business information systems among the Hungarian enterprises and analyses the following three key questions: how the usage of business information systems influences a company's economic performance, how much is the expenditure for an individual company to develop its information technology infrastructure and finally, to what extent information technology is considered important as a functional area within the organization of a company.

The aim of the research presented in this paper was to explore the current situation of Hungarian enterprises in terms of using business information systems, gaining a more thorough insight into the background of the decisions made on introducing such information systems together with the possible problems related to their introduction and further usage.

## II. THE AIM AND THE CONCEPT OF THE RESEARCH

The review of the relevant literature on the subject made it possible to identify the most important points of the research. Based on these, the main objectives as well as the concept of the research were formulated. The research objectives are the following:
- to present the background of the decisions related to the introduction of business information systems, along with the problems encountered in the phase of their introduction,
- to analyze the usage patterns of business information systems,
- to reveal the connection between using business information systems and the operational effectiveness or profitability of companies.

Based on the aims presented above, the following research concept was determined:
- First, the major issues related to the introduction of business information systems were analyzed. It was surveyed whether the companies taking part in the study used any sort of business information systems, and if not, what causes or conditions prevented them from introducing them. In the case of companies applying business information systems, the causes of introducing such systems, the information sources for selecting the appropriate systems, and the criteria for selecting them were also investigated. It was also examined whether companies had made calculations on the costs of introducing business information systems before making decisions on them, and if so, what aspects had been taken into account during the calculation. The problems occurring in the phase of implementation were identified.
- After that, the usage patterns of business information systems at companies were examined. The main points of the relevant analysis were the given company's information technology infrastructure, its Internet usage habits, and its appearance on the Internet. Here, the types of the applied business information systems and their areas of use were also presented, then the forms of information technology functions, human resource issues and the main points of IT strategy were covered.
- In the closing part of the analysis, the impacts of using business information systems on the operational effectiveness of companies were examined. It was investigated whether the introduction of such systems had an influence on the performance and revenues of the company as well as the size of the targeted market and the changes on the demand side. Another point of the investigation was whether ensuring more efficient information flow and information management contributed to the reduction of the company's other costs. These factors, summarized in Figure 1, of course, cannot be quantified so easily; however, taking them into consideration can lead to making more firmly grounded decisions on the introduction of various business information systems.

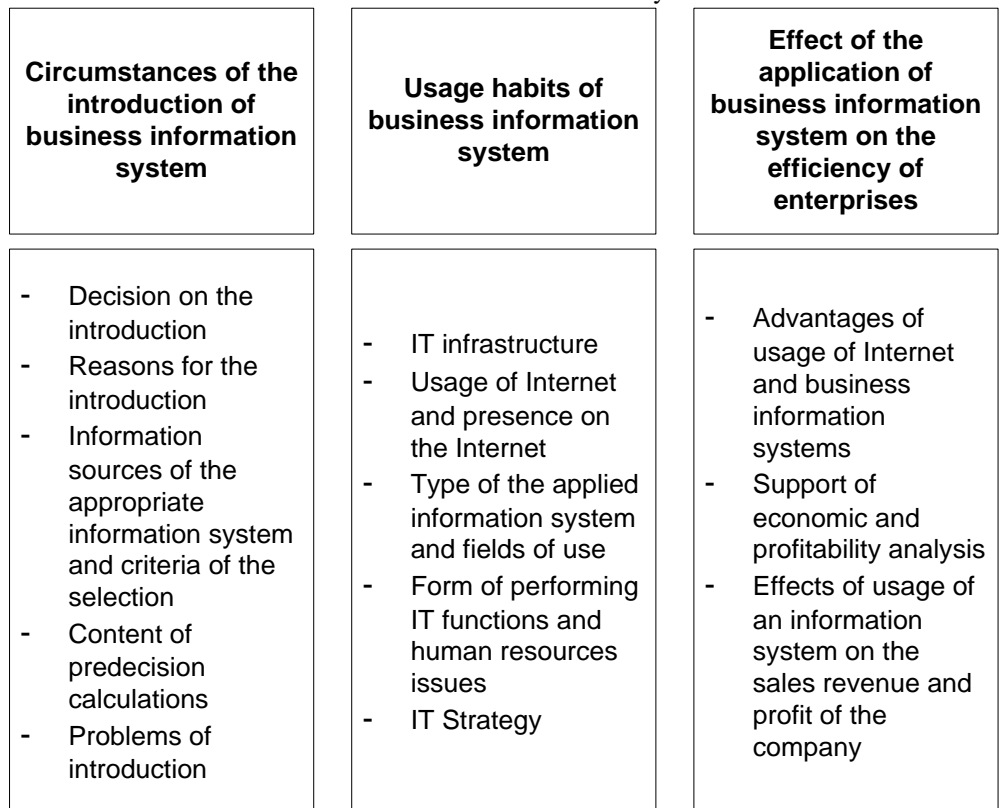

**Fig. 1 Conceptual model of the primary research**

### III. THE RESEARCH METHOD

The empirical survey was carried out using a written questionnaire. In the phase of compiling the individual questions of the survey, the main results of the previously conducted empirical surveys on the subject were also taken into account.

The questionnaire was divided into five major parts. The first part included some basic questions about the companies' background (such as their location, fields of operation, number of employees etc.), then questions related to the responding company's information technology infrastructure followed. In the third part of the questionnaire, the emphasis was put on questions enquiring about the Internet-using habits of the companies; the fourth group of questions was aimed at enquiring about the usage patterns of business information systems, making it the most detailed part of the questionnaire. The closing part contained questions about the IT-skilled human resources employed by the responding companies. The questionnaire was sent out to several hundreds of companies, The Hungarian survey was conducted both in a paper-based format and online with the assistance of the software application called Evasys. For evaluating data and presenting the results of the survey, the statistical software packages Excel 2007 and SPSS 19.0 were applied.

The 21% of the Hungarian responder companies are micro-sized, 29% are small-sized, 29% are medium-sized enterprise and 21% are corporations.

### IV. THE IMPACTS OF USING BUSINESS INFORMATION SYSTEMS ON OPERATIONAL EFFECTIVENESS

The analysis of the impacts of using business information systems covered the following three areas:
- The advantages originating from the use of the Internet and several business information systems,
- The support given to profitability calculations,
- The impact of applying business information systems on the company's turnover and overall performance.

Business expectations regarding the introduction of business information systems were investigated according to company size categories (see Fig. 2). The results clearly showed that according to the majority of the respondents the information supply for decision-makers improved. In contrast, business information systems as a means of improving competitiveness were only regarded by corporations, this aspect was only around the average in the other company size categories. Corporations also had the greatest expectations in terms of decreasing the time needed for making decisions and the improvement of internal communication, whereas microenterprises only moderately agreed with these statements. Corporations and medium-sized companies achieved an almost similar rate (4.04 and 4.35) when they were asked to evaluate using business information systems as an essential condition for keeping up with their competitors. Positive changes in the quality of the relationship with customers and suppliers were least expected by microenterprises (reaching a value of 3.42) while respondents representing the other size categories gave this aspect an average value. Only average values were given in every size category to the statement that costs could be reduced by using information systems.

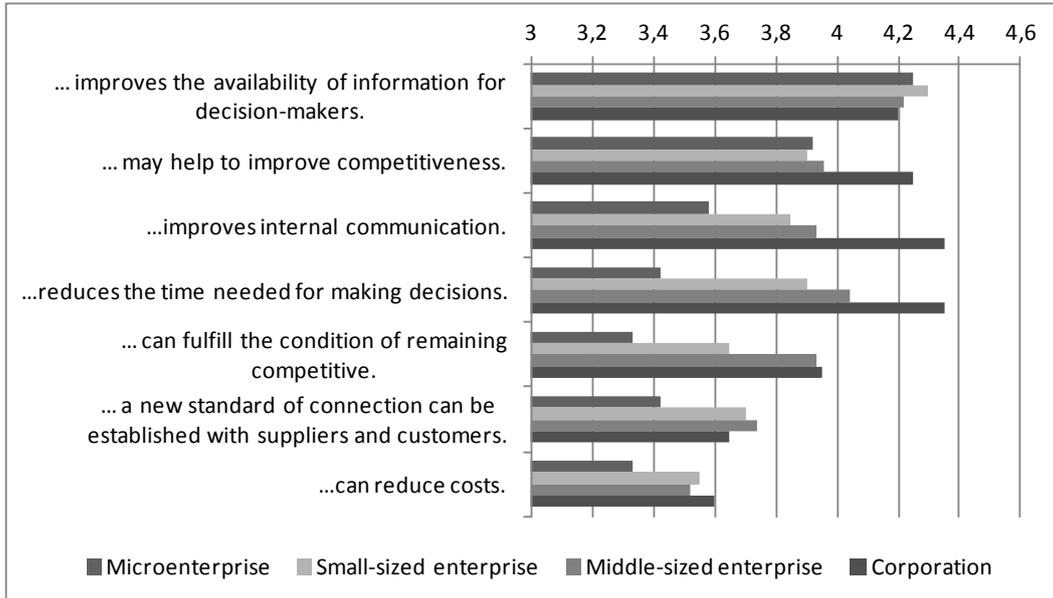

**Fig. 2 The evaluation of applying business information systems**

The respondents evaluated to what extent the usage of business information system provided assistance for their economic analyses. The obtained results show (Fig. 3.) that business information systems can be utilized well in controlling and reporting. There are differences in their judgement by size categories. Particularly corporations can take advantage of the support of business information systems mainly in the field of planning, plan-actual analysis and the exploration of cost reducing possibilities. As for the latter, the valuation of microenterprises is significantly behind of other size categories. Concerning the calculation of product profitability, small-sized enterprises slightly undervalue the support of business information systems compared to other company sizes. In the determination of product composition and answering to the questions of "buy or produce", the application of business information systems only provides a moderate support in all size categories.

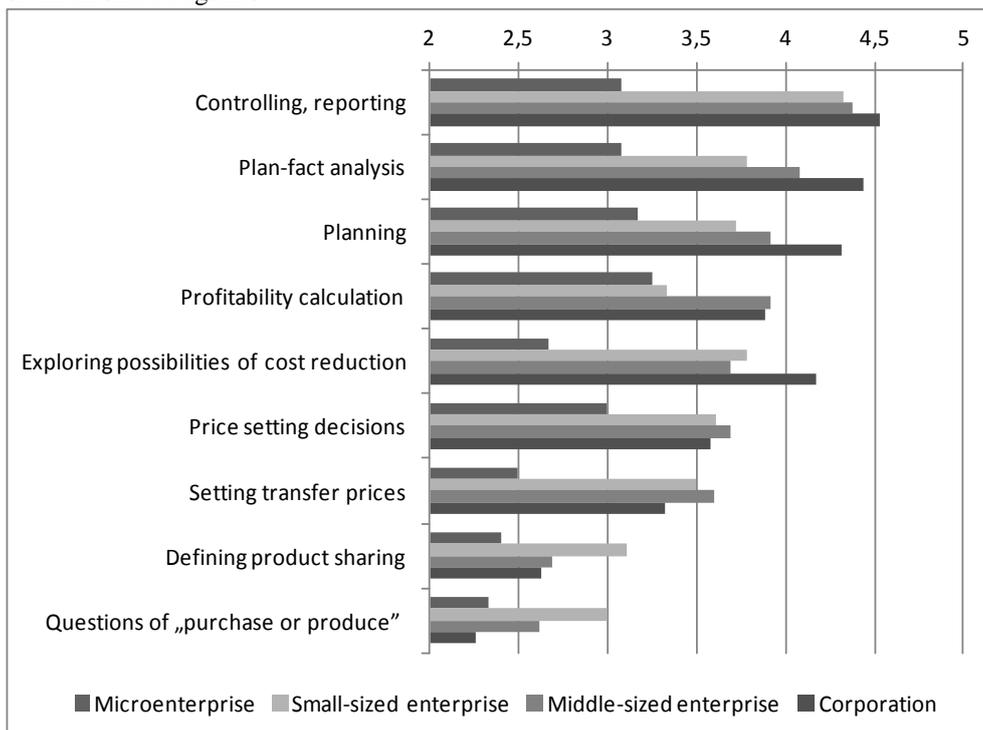

**Fig. 3 Evaluation of usage of business information systems in the field of economic analysis**

Based on these statements, I searched for a connection among the emerging costs of applying IT, the sales revenue and the profit before tax of the enterprises. The aim of the study was to demonstrate the positive impact of the application of business information systems and expenditures spent on IT on the sales revenue and the profit gained by the companies. The amount of IT expenditures is considered as an indicator for IT developments at the company. In my understanding, the more the company spends on IT investments and assets, the higher is the level of the development of its information technology. I carried out the examination by using the Pearson-correlation test, firstly aggregately to the companies of the sample and then separately by size categories as well. The analysis of the correlation was carried out among IT expenditures, earnings/profit before tax of the previous 4 years, aggregately and by separate size categories as well. First, I analyzed the relationship between the expenditures annually spent on IT and the sales revenue. Considering the sample as a whole, the correlation was not proven between the two variables, however examining the context by size categories, a positive, strongly significant relationship was proven in several cases. Except corporations (where correlation was not proven among the examined variables) it was proven in all size categories that IT expenditures emerged in the previous years (values in the year of 2007 and 2008) correlate with the sales revenue of some subsequent years (values in the year of 2009 and 2010). (The values of the Pearson correlation indicator were between 0.756-0.934 besides a 5% significance level.) Under the context, the companies that previously had spent more on IT investments earned a higher income in the subsequent years. Following that, I also carried out an analysis between the annual expenditures spent on IT and the profit before tax of the company. In this case, examining the sample by size categories as a whole and in part – also for the corporations – a strong, positive correlation was proven between the IT expenditures emerged in the previous years and profit before tax in the coming years. (The Pearson's correlation values were between 0.631 – 0.954 beside a 5% significance level, the deviation of the values was slightly higher than it was seen in the case of the sales revenue.) Naturally, the resulted context has to be treated with caution, as the sales revenue and the profit of the company is influenced by several other factors, the effects of which cannot be independently assessed. The different factors may strengthen or weaken each other's effect. In addition, I previously mentioned that the effects of the usage of business information systems on the sales revenue and the profit of the companies are difficult or impossible to be quantified. However, the existence of strong correlations still suggests that the positive effects of the usage of business information systems can be - if not numerically, but at least by the parallel movement of the variables - identified. It is also an important context that a strong relationship can be revealed among the IT expenditures of the previous years and the sales revenues and profits of the later years. This seems to confirm that the beneficial effects of IT investments are somewhat shifted and occur later in time.

## V. CONCLUSIONS

Nowadays the issue of information technology in business is moving into the centre of attention, which is also indicated by the fact that more and more companies, not accidentally, recognize its importance. Business information systems are not only fashionable – their application promotes more efficient operation of the company and also improves the supply of information to decision-makers; applying such systems can also play an important role in helping companies to put greater emphasis on information technology in order to gain a competitive advantage. My aim was to present the circumstances of the decisions made about the introduction of business information systems and problems emerging during the introduction as well as to analyze the usage habits of companies applying these systems, and to explore the relation between the application of business information systems and the operational effectiveness of the business. Based on the scientific literature, I worked out a conceptual model appropriate for the aims of the research, serving as a base both for the questionnaire and the analysis. The primary focus of the analysis was to explore the differences and similarities of the usage habits of business information system by size categories. Thus, the micro-, small and medium-sized enterprises as well as corporations were also presented in the sample. According to my observation, the correlation between the given factors could even be further strengthened by the application of complex statistical methods and by performing additional correlation assessments where the comparison should be carried out based on the main activity of the company rather than the size of the company, as I assume that the business scope of a company also determines the range of business information systems in use.

ACKNOWLEDGMENT

The described work was carried out as part of the TÁMOP-4.2.1.B-10/2/KONV-2010-0001 project in the framework of the New Hungarian Development Plan. The realization of this project is supported by the European Union, co-financed by the European Social Fund.

REFERENCES
[1]  M. Aranyossy, *Business Value of IT Investment: The Case of a Low Cost Airline's Website*. In: 20th Bled eConference eMergence: Merging and Emerging Technologies, Processes, and Institutions, June 4 - 6, 2007, Bled, Slovenia
[2]  B. Bencsik, *Az üzleti információs rendszerek használati szokásainak elemzése a vállalkozások körében (Analysis of the usage practice of business information systems among the enterprises)*, Szakdolgozat (MSC Thesis), Miskolc, 2011

[3]   L. Berényi, *Tudás(menedzsment) az irányítási szabványok sorai között. Knowledge(management) along the management standars*, GÉP LXIII:(6) pp. 17-20. 2012
[4]   E. Burt, and John A. Taylor, *Information and Communication Technologies: Reshaping Voluntary Organizations?*, Nonprofit Management and Leadership, Volume 11, Issue 2, pages 131–143, Winter 2000, 2003
[5]   P. Csala, A. Csetényi, and B. Tarlós, *Informatika alapjai (Basis of informatics)*, ComputerBooks, Budapest, 2003
[6]   L. Cser, and Z. Németh, *Gazdaságinformatikai alapok (Basis of economic informatics)*, Aula Kiadó, Budapest, 2007
[7]   G. B. Davis, and M. H. Olson, *Management information systems: Conceptual foundations, structure, and development.* New York: McGraw-Hill, 1985
[8]   I. Deák, P. Bodnár, and G. Gyurkó, *A gazdasági informatika alapjai (Basis of economic informatics)*, Perfekt Kiadó, Budapest, 2008
[9]   P. Dobay, *Vállalati információmenedzsment (Corporate information management)*, Nemzeti Tankönyvkiadó, Budapest, 1997
[10]  S. Floyd, and C. Wolf, *'Technology Strategy' In: Narayanan*, V.K. & O'Connor, G.C. (eds.) Encyclopedia of technology and innovation management. West Sussex: Wiley pp. 125-128. ISBN 1-4051-6049-7, 2010
[11]  Gy. Fülöp and G. I. Pelczné: *The SME-Sector Development Strategy in Hungary*, Global Management World Conference, Porto, Portugal, 2008
[12]  A. Gábor, *Üzleti informatika* (Business informatics), Aula Kiadó, Budapest, 2007
[13]  E. Garaj, *New Practice-Oriented Economic Knowledge and Learning Methods of Health Care Education*, Practice and Theory in Systems of Education 4:(2) pp. 15-22. 2009
[14]  C. Harland, *Supply Chain Management, Purchasing and Supply Management, Logistics, Vertical Integration, Materials Management and Supply Chain Dynamics*, Blackwell Encyclopedic Dictionary of Operations Management. UK: Blackwell, 1996
[15]  J. Hughes, *What is Supplier Relationship Management and Why Does it Matter?*, DILForientering, 2010
[16]  L. Kacsukné Bruckner and T. Kiss, *Bevezetés az üzleti informatikába (Introduction into business informatics)*. Akadémiai Kiadó, Budapest, 2007
[17]  S. Karajz, *Comparison of the Incentives to Innovate with Requirements, Taxes and Certificates*, University of Miskolc Innovation and Technology Transfer Centre, 3rd International Conference of PhD Students, pp. 345-352. 2/1., Engineering sciences, 2001
[18]  L. Kovacs, E. Gyöngyösi and L. Bednarik, *Development of classification module for automated question generation framework*, Teaching Mathematics and Computer Science 11:p. in press. 2013
[19]  P. Laudon, *Management Information Systems: Managing the Digital Firm*, Prentice Hall/CourseSmart, 2009
[20]  A. Nemeslaki, *Vállalati internetstratégia (Corporate Internet Strategy)*, Akadémiai Kiadó, Budapest, 2012
[21]  Zs. Majoros, *A Német és Magyar vállalkozások által alkalmazott Információs Rendszerek (Information technology systems used by the German and Hungarian enterprises)*, XXVII. microCAD International Scientific Conference, University of Miskolc, 21-22 March 2013
[22]  J. O'Brien, *Management Information Systems – Managing Information Technology in the Internetworked Enterprise*, Boston: Irwin McGraw-Hill, 1999
[23]  L.-D. Radu, *New dimensions of using ICTs in economics activities of organizations: environmental effects*, http://www.conferencedevelopments.com/files/Radux.pdf, 2011
[24]  M. Raffai, *Információrendszerek fejlesztése és menedzselése (Development and management of information systems)*. Novadat Kiadó, 2003
[25]  P Sasvari, *A Conceptual Framework for Definition of the Correlation Between Company Size Categories and the Proliferation of Business Information Systems in Hungary*, Theory, Methodology, Practice, Club of Economics in Miskolc, Volume 8: 2012, P 60-67, 2012
[26]  R. Shaw, *Computer Aided Marketing and Selling*, Rbhp Trade Group, ISBN 978-0750617079, 1991